# Role of rare-earth size on the structural, electronic and magnetic properties of $R_2$NiMnO$_6$ double perovskites

Mohd Nasir[1], Sunil Kumar[2], N. Patra[3], D. Bhattacharya[3], S.N. Jha[3], Dharma R. Basaula[4], Subhash Bhatt[4], Mahmud Khan[4], Shun Wein Liu[5], Sajal Biring[5], Somaditya Sen[1,5]

[1]Department of Physics, Indian Institute of Technology Indore, Indore, 453552, India
[2]Metallurgy Engineering and Materials Science, Indian Institute of Technology Indore, Indore, 453552, India
[3]Atomic & Molecular Physics Division, Bhabha Atomic Research Centre, Mumbai, 400085, India
[4]Department of Physics, Miami University, Oxford, Ohio 45056, USA
[5]Department of Electrical Engineering, Ming Chi University of Technology, New Taipei, 24301, Taiwan

The $R_2$NiMnO$_6$ ($R$ = La, Pr, Nd, Sm, Gd, Tb, Dy, Y, and Ho) double perovskites, prepared by sol-gel assisted combustion route, have been systematically investigated using powder x-ray diffraction, Raman spectroscopy, ultraviolet–visible spectroscopy, magnetization, and synchrotron based x-ray absorption spectroscopy measurements. All compounds in the family crystallize in the monoclinic structure (P2$_1$/$n$ space group) and the monoclinic distortion enhances with decreasing $R^{3+}$ radii. The magnetic ordering temperature ($T_C$) decreases from 270 K for La$_2$NiMnO$_6$ to 80 K for Ho$_2$NiMnO$_6$ as the $R^{3+}$ radii ($r_R^{3+}$) decrease from 1.16 Å (La$^{3+}$) to 1.02 Å (Ho$^{3+}$). An additional anomaly is observed in Nd$_2$NiMnO$_6$, Sm$_2$NiMnO$_6$, Tb$_2$NiMnO$_6$, and Dy$_2$NiMnO$_6$ at lower temperatures, which originates from the 3$d$-4$f$ coupling between Mn-Ni and Nd$^{3+}$/Sm$^{3+}$/Tb$^{3+}$/Dy$^{3+}$ magnetic moments. Further, high saturation magnetization is achieved for all samples indicating that they are atomically ordered and have less antisite disorders. Upon decreasing the size of $R^{3+}$, the local structure shows an expansion of NiO$_6$ octahedra and almost unchanged MnO$_6$ octahedra. X-ray-absorption near-edge spectroscopy reveals that majority of Ni and Mn ions are in +2 and +4 valence states in all the samples. Raman spectra of RNMO show a softening of phonon modes resulting in the elongation of Ni/Mn-O bond length. Finally, a correlation between lattice parameters, structural distortion, octahedral tilting, superexchange angle, and electronic band gap, Curie temperature, and the rare-earth ionic radius is established.

# I. INTRODUCTION

Double Perovskites $R_2BB'O_6$ ($R$ is a rare earth ion; $B$ and $B'$ are transition metals) have resurgence of scientific interest due to their wealthy physical properties and prospective of technological utilization[1-8]. Multifunctional properties of these perovskites include colossal magnetoresistance, magneto-capacitance, and multiferrocity. All such properties in $R_2BB'O_6$ compounds result from spin phonon coupling, which strongly depends on the $R$ and $B/B'$-site ordering[6-10]. Due to the observance of magneto-capacitance in $La_2NiMnO_6$ (LNMO)[11], rare-earth based double perovskites have been intensively investigated in the last decade. LNMO is a potential ferromagnetic semiconductor with coupled magnetic and electrical properties near room temperature ($T_C$~280 K)[11, 12].

Ferromagnetic semiconductor oxides, $R_2NiMnO_6$ (RNMO), are potential candidates due to their rich Physics and promising role in next generation spintronic industry. They can be widely utilized for fabricating novel magnetic, electronic, and multiferroic superlattices/systems because their properties can be tailored by changing the radii ($r_R^{3+}$) of the $R^{3+}$ ion as well as by applying external pressure[13]. Depending on the crystallographic site occupancies of Ni and Mn cations, these systems may adopt two different crystal structures. A random arrangement of Ni and Mn ions leads to an orthorhombic crystal structure with *Pbnm* space group, while an alternating periodic arrangement of Ni and Mn cations results in the monoclinic symmetry belonging to the P2$_1$/$n$ space group[14-16]. The $R_2NiMnO_6$ double perovskites exhibit high magnetic ordering temperatures[13]. The magnetic ordering originates from the virtual hopping of electrons between the half-filled orbital of $Ni^{2+}$ and the vacant orbital of $Mn^{4+}$ cations[17]. The superexchange interactions between the Ni and Mn ions, which control the magnetic ordering in RNMO, is governed by Goodenough-Kanamori rules [16, 18-20]. The magnetic interactions in $R_2NiMnO_6$ are also strongly influenced by $r_R^{3+}$. Lattice distribution of Mn and Ni ions will determine the superexchange interaction and hence the crystal structure. On decreasing $r_R^{3+}$, the ferromagnetic transition temperature decreases gradually, while the structure is distorted. Additionally, small antiferromagnetic clusters can be formed by less probable +3 oxidation states of Ni/Mn, antisite diorders, and antiphase boundaries leading to $Ni^{2+}$-$Ni^{2+}$ or $Mn^{4+}$-$Mn^{4+}$ interactions, which often results in the spin-glass state and metamagnetic transitions. The magnetic exchange interaction is associated with the deviation in octahedral tilting, Ni/Mn-O bond lengths and Ni-O-Mn bond angles. This kind of mechanism is well established for rare-earth based perovskites such as $RCrO_3$ and $R_2CoMnO_6$[21, 22].

$R_2NiMnO_6$ ($R \neq$ La) compounds are relatively less investigated and their electronic and magnetic properties, such as the saturation magnetization, magnetic ordering temperature ($T_C$), and the electronic bandgap vary considerably among various reports[14-16, 23]. Different conclusions have been drawn regarding the probable mechanism that governs magnetism in these materials. The reason for these differences may be due to the various synthesis routes that significantly influence the crystallographic structures and the extent of the antisite disorder at the Ni/Mn sites. Also, local atomic structure around Ni and Mn is a key factor in determining various physical properties such as magnetic ordering and electronic band gap in manganites, [24, 25]. For RNMO, the local structure and oxidation states of Ni and Mn ions are not well established. The detailed knowledge of the oxidation and spin states of Mn and Ni is critical to understand the broad magnetic response of these systems. Therefore, short-range-order techniques such as x-ray absorption spectroscopy (XAS), which directly probes local atomic structure have been employed to understand the physics of the RNMO series[26].

To address all ambiguities regularly distributed, homogeneous and single phase material is needed so that the real mechanism of the $R^{3+}$ dependence of magnetization can be explained properly. Hence, nine members of $R_2NiMnO_6$ ($R$ = La, Pr, Nd, Sm, Gd, Tb, Dy, Y, and Ho) family were synthesized by the sol-gel assisted combustion method. Taking into account the interesting variety of structure, electronic properties, and competing magnetic interactions in this series, a comprehensive study has been performed to ascertain the influence of the rare earth size on the structural and the magnetic ground states of these compounds. Such a comprehensive study on the entire RNMO series is a new addition to the ongoing research in the field.

## II. EXPERIMENTAL DETAILS

Polycrystalline $R_2NiMnO_6$ samples were prepared by a sol-gel assisted combustion method. For each $R_2NiMnO_6$ sample, stoichiometric amounts of the $R(NO_3)_3 \cdot 6H_2O$ ($R$ = La, Pr, Nd, Gd, Dy, Y), $Ni(NO_3)_2 \cdot 6H_2O$, and $Mn(NO_3)_2$ (50% w/w aqueous solution) were mixed together in doubly deionized water. $Sm_2O_3$, $Ho_2O_3$, and $Tb_4O_7$ were dissolved separately in doubly deionized with addition of few drops of nitric acid. The mixed solution was stirred vigorously for 40 minutes on a hot plate at ~50°C to ensure the through solubility. A gel former was prepared by mixing citric acid and ethylene glycol in the molar ratio 1:1. This gel former was poured into the clear solution. The resulting solution was heated at 85°C for 5 h on the hot plate with continuous stirring. After gel formation, the gels were burnt in open air at 100ºC to yield black fluffy powders. The obtained powders were ground using agate mortar and pestle and thereafter, calcined at 600ºC to get rid of trapped polymer and nitrates. The resulting powder was again ground and re-heated at 900 ºC for 12 h. After that, the powders were mixed with 5% PVA solution and pressed into pellets of diameter 10 mm and thickness 1.2 mm thickness using a uniaxial hydraulic press. The pellets were sintered first at 1150°C for 12 h to burn the binder (PVA) and finally at 1350 °C for 24 h to ensure compactness. The $R_2$NiMnO (RNMO) samples are labeled as LNMO ($La_2NiMnO_6$), PNMO ($Pr_2NiMnO_6$), NNMO ($Nd_2NiMnO_6$), SNMO ($Sm_2NiMnO_6$), GNMO ($Gd_2NiMnO_6$), TNMO ($Tb_2NiMnO_6$), DNMO ($Dy_2NiMnO_6$), HNMO ($Ho_2NiMnO_6$), and YNMO ($Y_2NiMnO_6$).

The phase identification and purity of these materials was investigated from powder X-ray diffraction data (XRD) collected using a Bruker D2 Phaser X-ray diffractometer with Cu $K_\alpha$ radiation. Rietveld refinement of XRD data using the *FullProf* software package [27, 28] was performed to extract the lattice parameters and bond angles. Temperature and field dependent magnetization (M-T and M-H) of the samples were examined on a Quantum Design Physical Property Measurement System (PPMS). The measurements were carried out in the temperature range of 4-300 K and at magnetic fields of up-to 5 T.

Room temperature x-ray absorption spectroscopy (XAS) measurements in fluorescence mode have been performed at the Energy Scanning EXAFS beamline BL-09 at Indus-2, RRCAT Indore, India [29, 30] for measuring the Ni (8333 eV) and Mn (6539 eV) $K$-edges of RNMO. The electron storage ring was operated at 2.5 GeV and 300 mA while the beamline works in the energy range 4-25 keV. The energy resolution of the beamline($\Delta E/E$) was about $10^{-4}$. In fluorescence mode, the sample was positioned at 45° to the incident X-ray beam. An ionization chamber measures the incident beam ($I_0$) while the flourosecnce signal ($I_f$) is measured using a Si drift detector placed at 90° to the incident X-ray beam. X-ray absorption co-efficient of the sample is determined by $\mu = I_f/ I_0$ and the spectrum was obtained as a function of energy by scanning the monochromator over the specified range.

For energy calibration, standard metal foils of Ni and Mn were used. The structure EXAFS signal $\chi(E)$ was calculated from the absorption coefficient $\mu(E)$ defined as [50]:

$$\chi(E) = \frac{\mu(E) - \mu_0(E)}{\Delta \mu_0(E_0)}$$

where, $E_0$ is absorption edge energy, $\mu_0(E_0)$ is the bare atom background, and $\Delta\mu_0(E_0)$ is the step in the $\mu(E)$ value at the absorption edge (difference between the pre edge and the post edge energy spectra). Finally the energy dependent EXAFS function $\chi(E)$ was converted to wave number dependent function $\chi(k)$ by $k = \sqrt{2m(E - E_0)/\hbar^2}$. After being weighted by $k^2$ which is used to compensate for the damping of the EXAFS amplitude with increasing $k$, the $\chi(k)$ data was Fourier transformed into $r$ space, in the range of $0 \leq k \leq 10$ Å. The background reduction of the experimental EXAFS data has been done by using the Athena module. The theoretical EXAFS spectra were generated using the FEFF 6.0 code while the fitting of the experimental EXAFS data with the theoretical spectra have been carried out using the ARTEMIS code of IFEFFIT software packages[31].

# III. RESULTS AND DISCUSSION
## A. Crystal structure

Room temperature powder x-ray diffraction was performed to examine the structural aspects of $R_2NiMnO_6$ series. X-ray patterns established that all samples were single phase except $Tb_2NiMnO_6$, $Ho_2NiMnO_6$ and $Y_2NiMnO_6$. The major phase belongs to a monoclinic structure belonging to the P2$_1$/n space group [Fig. 1(a)]. A minor impurity peak (marked with*) appears at ~29° for $Tb_2NiMnO_6$, $Ho_2NiMnO_6$ and $Y_2NiMnO_6$ corresponding to $Tb_2O_3$, $Ho_2O_3$ and $Y_2O_3$ respectively. Note that the ionic radii reduce in the order $r_{La} > r_{Pr} > r_{Nd} > r_{Sm} > r_{Gd} > r_{Tb} > r_{Dy} > r_Y > r_{Ho}$. All the samples show additional superstructure reflections at ~26°. These reflections become more intense with reducing ionic radius [Fig.1(c)]. A long range ordering between $Ni^{2+}$ and $Mn^{4+}$ cations is possible in a homogeneous sample. The peak appearing at $2\theta \approx 20°$ is the indication of long range ordering of $Ni^{2+}$ and $Mn^{4+}$ cations in RNMO samples. However, this feature is masked in the LNMO and PNMO samples. A continuous increase in the intensity of the reflection at ~26° is observed with decreasing rare-earth size. Splitting of peaks ~33° becomes more pronounced with decreasing rare-earth size, $r_R^{3+}$ [Fig. 1 (b)]. This emphasizes that the samples with smaller ionic radii get more stabilized in the monoclinic symmetry.

Rietveld refinement was employed to estimate structural parameters for each composition. We have considered monoclinic LNMO structure with P2$_1$/n space group as the starting model for refinement[32]. The lattice parameters, '$a$', '$b$' and '$c$' of this monoclinic symmetry, are related to the cell parameter, '$a_0$' ≈ 3.8 Å, of ideal cubic perovskite, as $a \approx b \approx \sqrt{2}a_0$, and $c \approx 2a_0$. The general 4$e$ ($x$, $y$, $z$) sites are occupied by $R$ ions and three dissimilar oxygen atoms (O1, O2 and O3). Ni and Mn atoms have two independent crystallographic positions, 2$d$ (0, 1/2, 0) and 2$c$ (1/2, 0, 0), respectively. The background was modeled using linear interpolation between a set background points with refineable heights. Bragg's reflections were modeled using Thompson-Cox-Hastings pseudo-Voigt Axial divergence asymmetry. Initially, scale factor and cell parameters were refined followed by profile and FWHM parameters. Once a proper match is achieved in the profiles, positional and thermal coordinates were refined. The oxygen atoms were not refined and were fixed to their stoichiometricity. All refinements reached a satisfactory fit. Representative refinement profiles for GNMO and DNMO are presented in Figure 2 and the representative refined crystal and local structure of $NiO_6$ and $MnO_6$ octahedra for DNMO are shown in Figure 3. In an ordered unit cell, the corner-shared $NiO_6$ and $MnO_6$ appear alternately illustrating two sublattices.

Figure 4 shows the evolution of structural parameters with $R^{3+}$ ionic radii as taken from Shannon for 8-fold oxygen coordination[33]. Figure 4(d) shows that Ni-O-Mn bond angles decreases almost linearly with $r_R^{3+}$. Angles decrease from 165.76° for LNMO to 144.02° for HNMO. The unit cell parameters $a$, $c$, and cell volume decreases with decreasing $r_R^{3+}$ from La to Ho [Figure 4 (a) and (c)]. However, variation in $b$ is small as compared to $a$ and $c$ parameters. This results from the tilting pattern of the type $a^-a^-c^+$ of $NiO_6$ and $MnO_6$ octahedra in $P2_1/n$ double perovskites[34]. In such kind of scheme, the distortion resulting from the reduction in $r_R^{3+}$ leaves b nearly invariant. It should be remarked that $c/\sqrt{2}$ lies between $a$ and $b$ for full RNMO series. The monoclinic distortion increases ($\beta$ deviates further from 90°) with reducing $R$-site cation radii [Fig. 4(b)]. The pseudocharacter of the cell reduces as the distortion of the perovskite structure enhances.

The tolerance factor ($t$) which is indicative of the stability of the structures, can be defined for $R_2NiMnO_6$ as,

$$t = \frac{r_R + r_O}{\sqrt{2}[\frac{r_{Ni}+r_{Mn}}{2} + r_O]}$$

where $r_R$, $r_{Ni}$, $r_{Mn}$, and $r_O$ are the effective ionic radii of $R$, Ni, Mn and O ions respectively [35]. The tolerance factor, $t$, strongly influences the magnetic and the electronic properties. With reducing $r_R^{3+}$, the value of $t$ decreases from 0.920 for LNMO to 0.849 for HNMO [Fig. 4(b)]. Tolerance factor is linked with the size mismatch between the $R$, $B$, and $B'$ cations and results in tilting of the octahedra with reduction in rear-earth size, $r_R^{3+}$. The tilting angle ($\varphi$) between the $NiO_6$ and $MnO_6$ octahedral can be calculated as $\varphi = [180 − (\angle Ni−O−Mn)]/2$. The tilting angle progressively increases with decrease in size of the rare earth, from 7.1° for LNMO to 17.9° for HNMO [Figure 4(d)]. This result emphasizes that cell distortion increases as the size decreases. Further, the deviation in $\beta$ from 90° is mainly due to the tilting of the octahedra. Such variations result in decrease of superexchange angle i.e. $Ni^{2+}$-$O^{2-}$-$Mn^{4+}$ angle.

## B. Vibrational properties

Raman spectroscopy is a robust technique to examine the crystal structure, cation disorder, local/dynamical lattice distortion, spin-phonon coupling and impurity phases present in the samples. The distortions from cubic $Fm\bar{3}m$ lattice results in low symmetry $R_2NiMnO_6$ ordered double perovskites. The vibrational modes arise from the (Ni/Mn)$O_6$ octahedra and $R$-O bonds. Room temperature Raman spectra for $R_2NiMnO_6$ samples were measured with excitation wavelength of 478 nm [Figure 5(a)]. For clarity, the Raman spectra are demonstrated in stacking configuration with upward base line shifts with decreasing rare-earth ionic radii, from La to Ho. The spectra exhibit two broad modes M1(~525 cm$^{-1}$) and M2 (~670 cm$^{-1}$). This is consistent with previous reports for ordered monoclinic LNMO, YNMO, RNMO and RCMO compounds[36-38]. Lattice dynamics calculations yield stretching "breathing" vibrations of $MnO_6$ and $NiO_6$ octahedra[39, 40], at ~ 670 cm$^{-1}$[41]. On the other hand, mixed modes of both antistretching and bending motions group at ~525 cm$^{-1}$. It has been critically discussed in literature, polarized Raman spectroscopy can distinguish between disordered and ordered phases with respect to the symmetry of the stretching mode. A broad band (1000-1450 cm$^{-1}$) associated with multiphonon scattering is also observed [Figure 5(c)]. Modes M3 (~1068 cm$^{-1}$) and M5 (~1340 cm$^{-1}$) are overtones of fundamental stretching modes M1 and M2, respectively and the mode M4 (~1206 cm$^{-1}$) is a combination of modes M1 and M2. All these three modes are relatively weak and broad. LNMO thin films were reported to exhibit such high frequency modes[42]. Due to the ordering of B/B' cations, phonons are not restricted to originate from a point near the Brillouin zone centre and can originate from anywhere within the Brillouin zone. This accounts for the second-order multiphonon process in the ordered double perovskites. Appearance of high frequency modes in RNMO samples hints that samples are doubly ordered. The first-order strong modes perceived near 525 cm$^{-1}$ and 670 cm$^{-1}$ are fairly asymmetric. Various factors are responsible for broadness and asymmetry in these modes. Firstly, this may be due to incomplete ordering of Ni and Mn sites and the various Ni/Mn-O vibrations have nearly same frequency so that unresolved components from (Ni/Mn)$O_6$ octahdra contribute to the band envelope. Secondly, domains with different ordering degrees and deviation from stable oxidation state of Ni and Mn contribute to the Ni/Mn-O stretching bands[38].

The replacement of La by smaller size $R^{3+}$ in $R_2NiMnO_6$ influences the position, symmetric nature and intensity of Raman peaks. Therefore, Raman spectra are deconvoluted to a sum of Lorentzians line profiles in the frequency range 450-825 cm$^{-1}$, which yields three peaks at ~ 494, 513 and 653 cm$^{-1}$ for samples with smaller rare-earth size. Notice that in samples with smaller $r_R^{3+}$, i.e., higher mass number, Raman band at ~ 520 cm$^{-1}$ is much more split than in La$_2$NiMnO$_6$. This effect can be ascribed to the higher monoclinic deformation of $R_2NiMnO_6$ with lower R size as compared to La$_2$NiMnO$_6$. This is quite consistent with monoclinc angle (β) as determined from XRD, which shows higher deviation as the rare-earth size is decreased. A representative fitted spectrum has been presented for GNMO in Figure 6(a). It can be noticed that this mode at ~ 520cm$^{-1}$ show clear softening with R-site ion changing from La to Ho [Figure 6(a)]. This is due to the increase of mass number and decrease of size of the R-site element from La to Ho. The evolution of the peak frequencies with $r_R^{3+}$ is illustrated in Figure 6(b) for the most prominent mode (670 cm$^{-1}$). The phenomena of phonon softening is ascribed to the increase in the average Ni/Mn–O bond length as smaller ionic substitution results in Ni-O bond length elongation. This nature can be explained within the harmonic approximation. A solid may be considered to be composed of atoms joined by springs having spring constant, k. In this approximation, the mode frequency (ω) is directly proportional to the square root of k, i.e, ω∝√k. The reduction in $r_R^{3+}$ results in decreasing lattice parameters and increasing bond lengths. Therefore spring constant is reduced thereby decreasing frequency with reducing $r_R^{3+}$.

Presence of weak stretching modes, in the low frequency regime, is shown by up arrows [Fig. 5(b)]. These modes are an indication of monoclinic P2$_1$/n symmetry. Hence, all samples are of monoclinic P2$_1$/n symmetry in consistence with structural properties analyzed by XRD. The low frequency vibrations are attributed to coupled (Ni/Mn)O$_6$ tilting vibrations and R-O stretching. These modes are activated by the monoclinic distortions and are more sensitive to changing rare-earth ionic radius. The most intense Raman mode of $R_2O_3$ at around 377 cm$^{-1}$ is not observed[43], confirming the phase purity of the samples. The main vibrational modes assigned to A$_g$ and B$_g$ are at 670 cm$^{-1}$ and 525 cm$^{-1}$, respectively[39, 44].

## C. Electronic properties: XANES analyses

XAS studies can reflect electronic properties and local site symmetry around an absorbing atom. This powerful tool has hence been employed to examine the charge state and the local environment of the constituent elements of $R_2NiMnO_6$ series. The room temperature Mn K-edge XANES spectra of RNMO are displayed in Fig. 7(a). The spectra of Mn foil, MnO,

$Mn_2O_3$, and $MnO_2$ are also shown as a reference for Mn $K$-edge. No striking difference is seen in the Mn $K$-edges spectral lines of RNMO samples. Spectra seem to be very similar to each other, and to the other half manganites[25, 45] having similar crystal structures. The spectra are described by two resonance peaks, marked as B (main resonance) near the edge and C (second resonance) in the post-edge region. The main resonance peak B also referred as white line is attributed to Mn1$s$→ Mn4$p$ transitions. They also exhibit a complex pre-edge structure A1 and A2 [inset (ii); Figure 7(a)] in the energy range 6537-6552 eV due to Mn3$d$-O2$p$ hybridisation. The second resonance peak arises from multiple scattering contributions of $MnO_6$ octahedra surrounded by La atoms. Spectra are nearly identical implying similar $MnO_6$ octahedral formation for all the RMNO samples. The subtle changes in intensity of the main resonance may imply different levels of distortion with change in the rare earth element. No difference is observed in the energy position of the absorption edge in RNMO spectra with change in rare-earth ionic size implying invariant valence state of Mn [inset (i); Figure 7(a)]. It is observed that absorption energies of these samples are close to standard $MnO_2$ spectrum. This reveals that Mn ions are predominantly in +4 oxidation state in all the samples. The complicated pre-edge structure clearly shows two peaks labeled as $A_1$(~6539 eV) and $A_2$(~6541 eV) having partial 3$d$ weight. These are present for the entire series. Such features are a consequence of mixing of quadrupole (1$s$→3$d$) and hitherto forbidden dipole (1$s$→3$d$) transitions due to hybridized 3$d$ and 4$p$ states. $A_1$ is attributed to transitions to majority $e_g$ levels whereas, feature $A_2$ corresponds to minority $e_g$ and $t_{2g}$ levels[46]. No significant change was observed in either the intensity or the energy position of the pre-edge structures for the RNMO samples. It means that Mn 3$d$ states show same degree of hybridization with the O 2$p$ with varying $R^{3+}$. In total, it is quite striking to realize that irrespective of the rare earth element, the $MnO_6$ octahedra do not change remarkably.

Figure 7(b) shows normalized XANES spectra at the Ni $K$-edge for $R_2NiMnO_6$ samples. Ni foil and NiO spectra are also shown as a standard. Similar to the Mn $K$-edge, the Ni $K$-edge spectra reveals two resonances, B' and C'. An additional shoulder D' is observed for some RMNO between B' and C'. The appearance of the shoulder D' can be related to changes in the local atomic environment around Ni. The pre-edge region (A') of the Ni K-edge data are not as intense as Mn $K$-pre-edges. Strength of $A_1$' peak (transitions to majority $e_g$ levels) is extremely poor and negligible in most RNMO samples. Only in case of TNMO a considerable $A_1$' peak is observed [Inset (ii); Figure 7(b)]. The $A_2$' (transitions to minority $e_g$ and $t_{2g}$ levels) feature is more pronounced compared to the $A_1$' in all RMNO samples.

Contrary to Mn, Ni pre-edge features show some change across the RNMO series which might be due to difference in the extent of hybridization between Ni-$3d$ and O-$2p$ states. The white line intensity is different for all samples. The absorption edge energies of Ni $K$-edges for RNMO samples are closer to edge energies of Ni $K$-edge for NiO [inset (ii); Fig. 7(b)]. This emphasizes that Ni is present in majority +2 valence states. A line broadening of the spectral features is observed with decreasing rare-earth size which can be attributed to the variation in $3d$ character, covalent nature of the ground state, and the symmetry across absorbing atoms. The broadening increases with decreasing size which causes Ni-O bond to expand [inset (iii); Fig. 7(b)]. It should be noted that Mn spectral lines show lesser broadening as compared Ni spectral ones. Thus the above XANES measurements show that Mn and Ni ions are primarily in +4 and +2 valence states, respectively.

Note that the ionic radii of $Mn^{4+}$(VI) is o.67Å while that of $Ni^{2+}$(VI) is 0.83Å. A homogeneous solution with such different size of octahedra is only possible when a proper alternate arrangement of $MnO_6$ and $NiO_6$ octahedra is possible hinting at a charge ordered state.

## D. Local structure: EXAFS analysis

To probe the local environment around Mn/Ni atoms, EXAFS has been performed on all samples. EXAFS analysis allows to monitor the changes in the (Mn/Ni)$O_6$ octahedra and Mn/Ni-O bond lengths.

Fourier transform (FT)of $k^2$-weighted Mn $K$-edge EXAFS spectra at room temperature was observed in the range 0-10 Å$^{-1}$ using a hanning window [Figure 8(a)]. A typical analyzed EXAFS signal exhibit similar oscillations for all samples [Figure 8(b)]. The EXAFS spectra exhibit a main peak at~1.35 Å for all RNMO samples. This peak corresponds to first Mn coordination shell to the six oxygen atoms surrounding Mn. The intensity and shapes of this peak remain almost invariant of the rare earth size. This indicates that oxygen coordination around Mn remains almost constant. Also, no significant change is noticed in peak positions of oxygen shell with smaller $r_R^{3+}$ indicating that the Mn-O bond lengths remain almost constant for the whole series. These observations hint at invariant $MnO_6$ octahedra in terms of size and shape. A set of weak and broad features appearing at ~2.08 and 2.69 Å correspond to Mn-La, Mn-Mn/Ni paths, which have not been considered in the fitting process. Their low intensities indicate a multi-path origin with some destructive interference. This observation is in agreement with the high monoclinic distortion of crystallographic structures of RNMO samples.

To extract further quantitative information of the Mn-O bonds, experimental EXAFS spectra were fitted [Figure 8(b)] with theoretical spectra using crystallographic data of $La_2NiMnO_6$. The coordination numbers were not refined and fixed to crystallographic compositions and the amplitude reduction factor ($s_0^2$) was fixed to 0.84 for whole RNMO series. Bond lengths, Debye-Waller ($\sigma^2$) factors, and threshold energy were varied during fitting. The nearest neighbour oxygen shell was analysed with one longer and one shorter bond lengths (2.15/1.94 Å) with 4 oxygen atoms in the shorter planar coordination while 2 more in the longer apical coordination, forming a distorted Jahn-Teller octahedral symmetry[47, 48]. Simulated best-fit parameters are summarized in Table I. No significant change is observed in Mn-O bond lengths suggesting that size and nature of the *R*-ion has little effect on the distortion or shape of $MnO_6$ octahedra.

The local environment of Ni cations has also been probed using Ni *K*-edge EXAFS spectra of all $R_2NiMnO_6$ samples. Figure 9(a) shows a typical $k^2$-weighted EXAFS experimental data. Fourier transforms (FT) using a Hanning window, between 0 Å$^{-1}$ to 10 Å$^{-1}$, are displayed in Figure 9(b). A prominent peak around 1.44 Å corresponds to the average Ni-O bond length. Coordination shells between 1.9 to 3.8 Å contribute to other scattering paths (Ni-O, Ni-Ni/Mn, and Mn-Mn). The first peak belonging to the Ni-O scattering paths reveals a notable change with decreasing $r_R^{3+}$. It progressively shifts towards higher values as $r_R^{3+}$ decreases from La to Ho, indicating expansion of Ni-O bondlengths [Figure 9(b)]. A decrease in peak intensity is noticed hinting to a larger distortion with reducing $r_R^{3+}$. Distortion may involve tilting of the $NiO_6$ octahedra. Structural analysis was performed in the range 0 - 4 Å in the *R*-space and fits are displayed in Figure 9(b). The fitting strategy is similar to the one used for Mn *K*-edge. The obtained structural parameters are tabulated in Table II for the first coordination shell. The high value of DW factor for HNMO sample, suggests a non-homogeneous distortion around the Ni atom. Ni-O bonds increase with reducing $r_R^{3+}$ [Table II]. Expansion of the apical axis, *i.e.* the longer Ni-O bond length, can be related to an increasing $NiO_6$ rotation. The calculated bond lengths of $Ni^{2+}$-$O^{2+}$/$Ni^{3+}$-$O^{2-}$ and $Mn^{3+}$-$O^{2-}$/$Mn^{4+}$-$O^{2-}$ using Shannon radii table are 2.04 Å/1.95 Å and 1.995 Å/1.88 Å respectively. From the EXAFS results we found Ni–O and Mn–O average bondlength ~ 2.04 Å and 1.95 Å respectively. With decreasing $r_R^{3+}$, Ni–O bondlength increases, indicating stabilization of the bivalence state of Ni ions.

Thus, EXAFS results reveal an increase of average Ni-O bondlength without significant changes of Mn–O bondlength with decreasing $r_R^{3+}$. Oxygen atoms shared by both $R$ and Ni atoms moves towards $R$ atoms, with decreasing $r_R^{3+}$. Distortion in the octahedra can be estimated from the reduction and increment of 1$^{st}$ co-ordination shell volume of the R–O and Ni–O shells respectively. There is no variation in the 1$^{st}$ co-ordination shell of the Mn atoms with decreasing $r_R^{3+}$. Also, NiO$_6$ octahedra are larger than MnO$_6$ ones. EXAFS, XANES, XRD and Raman spectroscopy results are in agreement with each other.

## D. Magnetic properties

Octahedral tilting, tolerance factor, lattice parameters and rare-earth size greatly influences the magnetic properties of a perovskite/double perovskite structure. The orbital overlap of B–O–B' bonds are the key players in this aspect.

Temperature dependent dc magnetization, M(T), of $R_2$NiMnO$_6$ samples, have been performed in the temperature range of 4-300 K under 100 Oe magnetic field. Samples were cooled from room temperature down to 4K in absence of magnetic field. After the application of a magnetic field at the lowest temperature, zero field cooling, ZFC, M(T) measurement was performed during heating of the sample from 4K to 300K at the same constant stable magnetic field (~100 Oe). Field cooled cooling, FCC, M(T) measurements were performed during cooling the sample in field following the ZFC measurement. The ZFC and FCC M(T) data for all the $R_2$NiMnO$_6$ samples are shown in Fig. 10, and the results are in good agreement with earlier reports[14, 16, 20, 49]. Below the ferromagnetic transition temperature, $T_C$, thermomagnetic irreversibility is observed for all samples in the ZFC and FC M(T) data. This behavior is typical for $R_2$NiMnO$_6$ compounds[20]. Bifurcation between FC and ZFC curves hints at competing magnetic interactions or spin geometric frustrations like spin glass or cluster glass at lower temperatures [50].

An additional anomaly is observed at lower temperatures for the NNMO, SNMO, TNMO, and DNMO samples [Fig. 10]. This low temperature anomaly is ascribed to coexisting FM (major phase) and AFM (minor phase) ordering caused by 3$d$-4$f$ exchange interactions between Mn-Ni and Tb$^{3+}$/Nd$^{3+}$/Dy$^{3+}$/Sm$^{3+}$ magnetic moments[16, 20, 51]. Interestingly, negative ZFC M(T) is noticed in the PNMO, NNMO, YNMO and HNMO samples. However, it is positive for all other samples. This is attributed to formation of the antiphase boundary generating from the antisite defects at Ni/Mn sites. In such cases, a microscopic domain structure is comprised of a number of ferromagnetic clusters and domains. A short-ranged ferromagnetic coupling exists between $Ni^{2+}$ and $Mn^{4+}$ ions and an antiferromagnetic coupling takes place in the antiphase boundary. Infact, a field of 100 Oe is insufficient in the ZFC mode to align the spins in the direction of applied magnetic field across antiphase boundary. Also, the antiparallel or canted spins get more stabilized as the temperature is decreased. Similar frustration effect was observed in $Sr_2NiReO_6$ and $La_{2-x}Bi_xCoMnO_6$ double perovskites[52, 53].

Curie-Weiss law defines the paramagnetic component above $T_C$: $\chi = C / (T - \theta_{Weiss})$, where $C$ is Curie constant and $\theta_{Weiss}$ is the paramagnetic Curie temperature (Weiss constant). Above $T_C$, a linear behavior in inverse susceptibilities has been observed for all compounds. Note that although DNMO, HNMO, and TNMO show some deviation from linearity of these plots of high temperature reciprocal susceptibility, the T-axis intercept still remains positive [Fig. 11(a)]. Hence, the paramagnetic temperatures are positive for all samples confirming the predominant magnetic interactions in $R_2NiMnO_6$ series are ferromagnetic in nature. The effective paramagnetic moment, $\mu_{eff}$, is obtained from the Curie-Weiss fit using the relation $\mu_{eff} = 2.828 \sqrt{C}$. The parameters $C$, $\theta_{Weiss}$ and $\mu_{eff}$ are documented in Table III. The values for $\theta_{Weiss}$ are larger than $T_C$ for LNMO, PNMO, NNMO, and SNMO and smaller for GNMO, TNMO, DNMO, YNMO and HNMO sample. The theoretical effective magnetic moments are calculated using the relation, $\mu_{eff}(calc) = [2\mu_B(R^{3+})^2 + \mu_B(Ni^{2+})^2 + \mu_B(Mn^{4+})^2]^{1/2}$.[16] The effective ground state paramagnetic moments, 3.58 $\mu_B$ for $Pr^{3+}$, 3.62 $\mu_B$ for $Nd^{3+}$, 0.85 $\mu_B$ for $Sm^{3+}$, 7.94 $\mu_B$ for $Gd^{3+}$, 9.72 $\mu_B$ for $Tb^{3+}$, 10.65 $\mu_B$ for $Dy^{3+}$, 10.60 $\mu_B$ for Ho, 3.87 $\mu_B$ for $Mn^{4+}$, and 2.83 $\mu_B$ for $Ni^{2+}$ have been used for calculations. The $\mu_{eff}(calc)$ are also provided in Table III. The $\mu_{eff}$ obtained from Curie-Weiss law are in agreement with calculated values for all RNMO samples. The $\mu_{eff}$ for LNMO is larger than the corresponding theoretical value, $\mu_{eff}(calc)$. A similar behavior had been reported and explained as an unusual paramagnetic state consisting of superparamagnetic clusters and domains, near ferromagnetic ordering temperature[17, 50, 54]. The ferromagnetic interactions exit within the clusters and domains while paramagnetic state is observed in between domains or clusters. This is responsible for large paramagnetic moment, $\mu_{eff}$. The larger value of $\mu_{eff}$ has also been reported for $La_2CoMnO_6$ double perovskites[55] and $LaMnO_{3+\delta}$[54] perovskites. Actually, a wider temperature range ~250-700K is required to get a better Curie-Weiss fitting. The $\mu_{eff}$ values for PNMO, NNMO, HNMO, and TNMO are very close to the reported ones for the ordered perovskites[15, 16, 35].

As shown in Fig. 11(b), the $T_C$ (estimated from the derivative of M(T) data) decreases with decreasing $r_R^{3+}$. This monotonic decrease of $T_C$ with decreasing $r_R^{3+}$ hints at correlation of magnetism with Ni/Mn-O bond length and the Ni-O-Mn bond angle. Mechanism of magnetism is mainly governed by superexchange process as per Goodenough−Kanamori rules. The superexchange coupling arises from the virtual hopping between half-filled Ni ($e_g^2$) and empty Mn ($e_g^0$) orbitals. This gives rise to a set of ferromagnetic ordering tailored by Ni/Mn-O distance and the Ni-O-Mn bond angle. X-ray diffraction revealed that <Ni–O–Mn> bond angles decrease with decreasing rare-earth size from 165º for LNMO to 144º for HNMO. This reduction in bond angles reduces the overlap between the orbitals, which in turn lowers the strength of superexchange interaction between the $Ni^{2+}$ and $Mn^{4+}$ elements. This is also realized from the reduction in tolerance factor from LNMO to HNMO[15]. Thus, Curie temperature decreases from 270 K for LNMO to 80 K for HNMO. The strong correlation found between $T_C$ and Ni–O–Mn bond angles and the rare-earth size, emphasizes that $T_C$ is actually being controlled by the nature and size of rare-earth element. Quantitative EXAFS analysis shows that average Mn/Ni-O bond distances increase with reduction in rare-earth size, thereby decreasing Mn/Ni-O covalent character. This decreasing covalency is also accountable for the decrease of $T_C$.

Isothermal field dependent magnetization hysteresis loops for $R_2NiMnO_6$ at 5 K [Fig. 12] reveal ferromagnetic nature for all samples (see inset of Fig. 12). The M-H curves do not saturate at 5T. The behavior of M-H curves depends on the magnetic nature of $R^{3+}$ ions. The magnetization increases continuously for GNMO, TNMO, DNMO, and HNMO samples up to field of 5 T. This might be arising from the polarization of the $Gd^{3+}/Tb^{3+}/Dy^{3+}/Ho^{3+}$ moments. The theoretical saturation moments ($M_s$), the measured ($M_{5T,5K}$) and corresponding extrapolated moment values ($M_{extrapolated}$) are listed in Table III. The $M_s$ for all samples has been calculated using, $M_s=[2g_J + 5.0]\mu_B$/f.u, where $g_J$ is the rare-earth statured moment and $5\mu_B$/f.u is saturated moment for $Ni^{2+}$ and $Mn^{4+}$ ions. The extrapolated moment values ($M_{extrapolated}$) are extracted from M *vs.* 1/H extrapolated to 1/H = 0. Note that the measured magnetic moments are smaller than the theoretical ones, yet these samples have higher magnetization than experimentally reported values[14, 23]. This indicates that these samples have better magnetic ordering and less antisite disorders. This is possible because these samples are of better crystalline nature and homogeneity. Although these samples exhibit relatively better crystalline properties, perfect lattice don't exist in them. Hence, a perfect collinear arrangement is not possible resulting in canted structures. Therefore comparison to a theoretical value is not justified as the $Ni^{2+}/Mn^{4+}$ spins, and $R^{3+}$ moments are not perfectly ordered. Also, antisite disorders largely influence magnetization which in turn depends on synthesis routes, sintering temperature and reaction time, *etc*. For example, in present study, LNMO sample has a saturation moment of 4.2 $\mu_B$/f.u., emphasizing a more ordered state than reported previously for LNMO with a saturation moment of 3.2 $\mu_B$/f.u[56]. The saturation moment of 5 $\mu_B$/f.u. is expected for a perfectly ordered LNMO system because $Ni^{2+}$ contributes two unpaired electrons and $Mn^{4+}$ contributes three unpaired ones. Thus, a net magnetic moment of 4.2$\mu_B$/f.u., with a reduction of 0.8 $\mu_B$/f.u. from the perfectly ordered moment of 5 $\mu_B$/f.u., suggests the presence of ∼8% antisite defects. Thus, we have less antiste defects in our all samples hinting at high degree of ordering, resulting in large magnetization. Besides, values of remnant magnetization ($M_r$) andcoercivity ($H_c$) for all samples measured in this study, which indicate the stability of ferromagnetic coupling, are significantly large.

### D. Electronic band gap

To study the *R*-dependence of optical properties of RNMO, the electronic band gap is obtained from the UV-Visible absorption spectra using Tauc relation[57] given by, $\alpha h\nu = A(h\nu - E_g)^n$, where, $h\nu$ is the incident photon energy, $\alpha$ is the absorption coefficient, A is a characteristic parameter independent of the photon energy, $E_g$ is the electronic band gap, and n is a dimensionless parameter with a value of 1/2 for direct-allowed transitions. The $(\alpha h\nu)^2$ *vs* $h\nu$ plot reveals a linear nature near the absorption edge [ Fig. 13]. The band gap energy is extracted by extrapolating the linear part of this curve with a straight line to $(\alpha h\nu)^2 = 0$ axis. Bandgap of RNMO increases with decreasing $r_R^{3+}$ from 1.42 eV for LNMO to 1.68 eV for YNMO [Fig. 14]. Theoretical studies support this result[58]. In a similar study on $R_2CoMnO_6$, such trend of increasing band gap with decreasing $r_R^{3+}$ has also been reported[59].

Omrane *et al.* reported O-2*p* states (-5 to 0 eV) below but *d*-states of La, Ni, and Mn above the Fermi level in the energy band structure of $La_2NiMnO_6$ and $La_2CoMnO_6$ perovskites[60]. Ma et. al attributed the valence band to the interaction of Ni 3*d* orbitals with the O 2*p* orbitals. On the other hand they pointed out that hybridization between Mn 3*d* and O 2*p* orbitals may be responsible for the bottom of the conduction band[61]. Hence bandgap, $E_g$, depends on the overlap of Ni/Mn and O orbitals. The B–O–B' bond exchange angle, θ, is a major factor which determines the degree of overlap and hence the binding energy in double perovskites[59]. The exchange angle, ∠Ni-O-Mn, decreases with reducing $r_R^{3+}$. Hence, orbital overlap decreases resulting in a larger band gap, but a weaker super-exchange interaction. The linear correlation between $E_g$, $T_C$ and ∠Ni-O-Mn indicates that the superexchange angle is possibly responsible for the increment in band gap [Fig. 14]. Experimental evidence[38] in these samples further strengthen the existing results as well as agrees with theoretical observations. Thus, a wide tunability of bandgap by *R*-site substitution in such perovskites can be achieved as the optical properties are influenced by crystal structures.

## IV. CONCLUSIONS

Sol-gel prepared homogeneous, single phase $R_2NiMnO_6$ perovskitescrystallize in a monoclinic structure having only P2$_1$/*n* space group. Lattice parameters decrease resulting in decrease of unit cell volume with decreases of rare-earth ionic size ($r_R^{3+}$). Structural distortions like octahedral tilting and monoclinic distortions become stronger with decreasing $r_R^{3+}$. To maintain orbital overlap between $O^{2-}$ and $R^{3+}$ ions, with reducing $r_R^{3+}$, octahedral tilting needs modification. Decrease in $r_R^{3+}$, is followed by decrease in superexchange angle (Ni-O-Mn) from 165º for LNMO to 144º for HNMO. This is followed by Ni(Mn)-O bond length elongation and phonon softening. EXAFS studies support such observations. XANES studies confirm $Ni^{2+}$ and $Mn^{4+}$ valence states in all $R_2NiMnO$ samples, irrespective of $r_R^{3+}$. Increasing distortion and $NiO_6$ octahedra tilting is confirmed from local structure analysis. Magnetic studies indicate *B*-site ordering in all the samples with a single ferromagnetic transition Curie temperature, $T_C$. With decreasing $r_R^{3+}$, $T_C$ reduces from~ 270 K to 80 K. Decrease in superexchange interaction between $Ni^{2+}$ and $Mn^{4+}$ ions with decreasing $r_R^{3+}$ may be responsible for such a variation. Participation of rare earth magnetic moments occurs at low temperature, as evident from the dc magnetization data of $Tb_2NiMnO_6$, $Nd_2NiMnO_6$, $Dy_2NiMnO_6$, and $Sm_2NiMnO_6$. This magnetic ordering originates from 3*d*-4*f* exchange interactions between Mn-Ni and $Tb^{3+}/Nd^{3+}/Dy^{3+}/Sm^{3+}$ magnetic moments. It is also observed that the electronic bandgap can be tuned by $r_R^{3+}$. Hence, magnetic and electronic properties of $R_2NiMnO_6$ compounds can be manipulated by $r_R^{3+}$. This makes these materials potential candidates for next generation spintronic applications.

### ACKNOWLEDGMENTS

Mohd. Nasir gratefully acknowledges financial support received from UGC, New Delhi under Maulana Azad National fellowship. Dr. Pankaj R. Sagdeo is acknowledged for helping in UV measurements. Dr. Sajal Biring acknowledges the financial support from Ministry of Science and Technology, Taiwan (MOST 105-2218-E-131-003 and 106-2221-E-131-027).

# Notes and references

# Figure Captions

Fig. 1. (a) X-ray diffraction patters of $R_2NiMnO_6$ ($R$= La, Pr, Nd, Sm, Gd, Tb, Dy, Y, and Ho). All the peaks were indexed to the monoclinic cell belonging to P2$_1$/$n$ space group. (b) Magnified view of peak ~ 33º. (c) Evolution of peaks ~ 26º with decreasing rare-earth size ($r_R^{3+}$).

Fig. 2. Rietveld refinement of X-ray diffraction patters of (a) Nd$_2$NiMnO$_6$ and (b)Dy$_2$NiMnO$_6$ samples. Cross symbols (black color): experimental data points; Solid line (red color): Simulated curve; vertical sticks (Braggs reflections); the bottom olive line is the difference between the experimental and the calculated patterns. The enlarged view of the fitted main peak for (c) DNMO and (d) NNMO.

Fig. 3.(a) Crystal structure and (b) local structure of NiO$_6$ and MnO$_6$ octahedra for Dy$_2$NiMnO$_6$ as generated from Vesta program.

Fig. 4. Evolution of the structural parameters of $R_2NiMnO_6$ double perovskite oxides as a function of the rare-earth size, $r_R^{3+}$. (a) cell parameters $a$, $b$, and $c$; (b) monoclinic angle $β$ and tolerance factor, $t$; and (b) Ni-O-Mn bond angles and tilting angle.

Fig. 5. (a) Typical unpolarized Raman spectra for $R_2NiMnO_6$ samples at 298 K; A magnified view of the (b) low-frequency and (c) high-frequency spectral features. For clarity, spectra have been stacked vertically. The dotted vertical line is a guide to the eye to show the shift of modes with $r_R^{3+}$.

Fig. 6. (a) The deconvoluted Raman spectra of GNMO using Lorenzian line shape function. The shaded areas confirm the presence of three modes, and (b) Softening of mode at ~ 670 cm$^{-1}$ with $r_R^{3+}$.

Fig. 7. Normalized (a) Mn *K*-edge and (b) Ni *K*-edge XANES spectra for RNMO compounds. Inset I: First derivative of XANES spectra and Inset II: Pre-edge region of the XANES spectra for same compounds. Inset III: magnified view of main resonance.

Fig. 8. (a) $k^2$-weighted EXAFS spectra at the Mn $K$-edge of RNMO samples and (b) Modulus of the Fourier transforms along with its fits (red lines) for the RNMO samples at the Mn $K$-edge.

Fig. 9. (a) $k^2$-weighted EXAFS spectra at the Ni $K$-edge of RNMO samples and (b) Modulus of the Fourier transforms together with its fits (red lines) for the RNMO samples at the Ni $K$-edge.

Fig. 10. Temperature dependence of the field cooled cooling (FCC) and zero field cooling (ZFC) dc magnetization for $R_2NiMnO_6$ double perovskites under 100 Oe applied magnetic field.

Fig. 11.(a) The inverse susceptibility for $R_2NiMnO_6$ and fits to the high temperature using Curie-Weiss law, and (b) Variation of the magnetic ordering temperature, $T_C$, of $R_2NiMnO_6$ as a function of rare earth size, $r_R^{3+}$.

Fig. 12. Magnetization *vs.* magnetic field isotherms for $R_2NiMnO_6$ double perovskites at 5 K temperature.

Fig. 13. UV-Visible absorption spectra for $R_2NiMnO_6$ double perovskites for direct transition $(\alpha h\nu)^2$ *vs* $h\nu$.

Fig. 14. Electronic band gap and Curie temperature, $T_C$ of $R_2NiMnO_6$ as a function of (a) superexchange angle ∠Ni-O-Mn and (b) ionic radii $r_R^{3+}$.

**TABLE I.** EXAFS parameters the oxygen coordination shell for RNMO samples at the Mn $K$-edge. C.N. is the coordination number, and R is the interatomic Mn-O distances. $\sigma^2$ are the absolute values for the Debye-Waller factors.

| Samples | Mn – O1 | | | Mn – O2 | | |
|---|---|---|---|---|---|---|
| | C.N | R(Å) | $\sigma^2$ (Å$^2$) | C.N | R(Å) | $\sigma^2$ (Å$^2$) |
| LNMO | 6 | 1.89(1) | 0.004(2) | | | |
| PNMO | 6 | 1.85(2) | 0.005(2) | | | |
| SNMO | 4 | 1.90(2) | 0.002(1) | 2 | 2.39(4) | 0.002(2) |
| GNMO | 4 | 1.89(2) | 0.002(2) | 2 | 2.00(4) | 0.010(9) |
| TNMO | 4 | 1.90(1) | 0.002(5) | 2 | 2.02(5) | 0.008(8) |
| DNMO | 4 | 1.89(1) | 0.002(3) | 2 | 2.04(1) | 0.011(9) |
| HNMO | 6 | 1.89(3) | 0.008(3 | | | |
| YNMO | 4 | 1.89(1) | 0.003(4) | 2 | 1.99(5) | 0.012(9) |

**TABLE II.** EXAFS parameters of the oxygen coordination shell for RNMO samples at the Ni *K*-edge. C.N. are the coordination numbers, and R are the interatomic Ni-O distances. $\sigma^2$ are the absolute values for the Debye-Waller factors.

| Samples | Ni – O1 | | | Ni – O2 | | |
|---|---|---|---|---|---|---|
| | C.N | R(Å) | $\sigma^2$ (Å$^2$) | C.N | R(Å) | $\sigma^2$ (Å$^2$) |
| LNMO | 4 | 1.97(1) | 0.005(1) | 2 | 2.02(4) | 0.005(2) |
| PNMO | 4 | 1.98(3) | 0.002(3 | 2 | 2.13(2 | 0.003(9 |
| NNMO | 4 | 2.01(2) | 0.003(2) | 2 | 2.06(2) | 0.008(9) |
| SNMO | 4 | 2.01(3) | 0.003(3) | 2 | 2.06(2) | 0.008(6) |
| GNMO | 4 | 2.02(1) | 0.004(2 | 2 | 2.09(5) | 0.008(6) |
| YNMO | 4 | 2.02(2) | 0.005(3 | 2 | 2.06(3) | 0.005(5) |
| DNMO | 4 | 2.03(2) | 0.004(3) | 2 | 2.09(3) | 0.007(4) |
| HNMO | 4 | 2.05(4) | 0.004(2 | 2 | 2.05(4) | 0.004(2) |
| YNMO | 4 | 2.00(2) | 0.003(2) | 2 | 2.13(5) | 0.010(9) |

**TABLE III.** Magnetic parameters for $R_2NiMnO_6$ double perovskites

| Sample | LNMO | PNMO | NNMO | SNMO | GNMO | TNMO | DNMO | YNMO | HNMO |
|---|---|---|---|---|---|---|---|---|---|
| $C$ (emu.k.mol$^{-1}$.Oe$^{-1}$) | 5.97 | 4.69 | 5.12 | 4.12 | 13.93 | 21.74 | 25.37 | 3.06 | 25.40 |
| $\theta_{Weiss}$ (K) | 273 | 218 | 204 | 174 | 101 | 60 | 42 | 104 | 31 |
| $\mu_{eff}(\mu_B/fu)$ | 6.910 | 6.124 | 6.400 | 5.744 | 10.556 | 13.187 | 14.245 | 4.95 | 14.254 |
| $\mu_{cal}(\mu_B/fu)$ | 5.25 | 6.97 | 7.04 | 4.94 | 12.23 | 14.26 | 15.80 | 5.25 | 15.47 |
| $T_C$ (K) | 270 | 210 | 195 | 160 | 130 | 110 | 95 | 79 | 80 |

| $M_r$ ($\mu_B$/f.u.) | 1.09 | 1.10 | 0.65 | 0.95 | 0.92 | 1.53 | 1.13 | 1.24 | 1.31 |
|---|---|---|---|---|---|---|---|---|---|
| $H_c$ (Oe) | 255 | 336 | 688 | 311 | 162 | 287 | 358 | 362 | 429 |
| $M_{5K, 5T}$ ($\mu_B$/f.u.) | 4.24 | 4.92 | 5.65 | 5.05 | 16.21 | 13.26 | 13.42 | 4.85 | 13.23 |
| $M_{extrapolated}$ ($\mu_B$/f.u.) | 4.39 | 5.48 | 7.14 | 5.21 | 21.37 | 15.38 | 16.28 | 4.97 | 15.52 |
| $M_s$ ($\mu_B$/f.u.) | 5.0 | 11.4 | 11.6 | 6.4 | 19.0 | 23.0 | 25.0 | 5.0 | 25.0 |

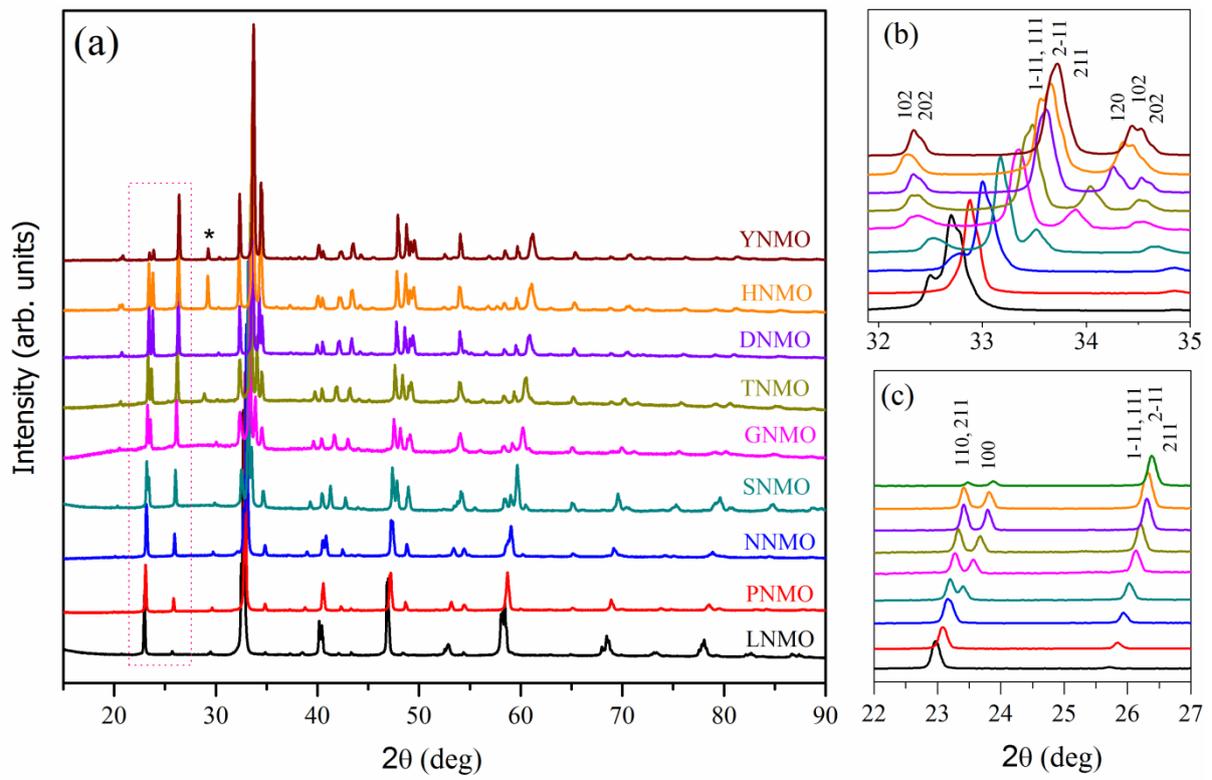

Fig. 1.

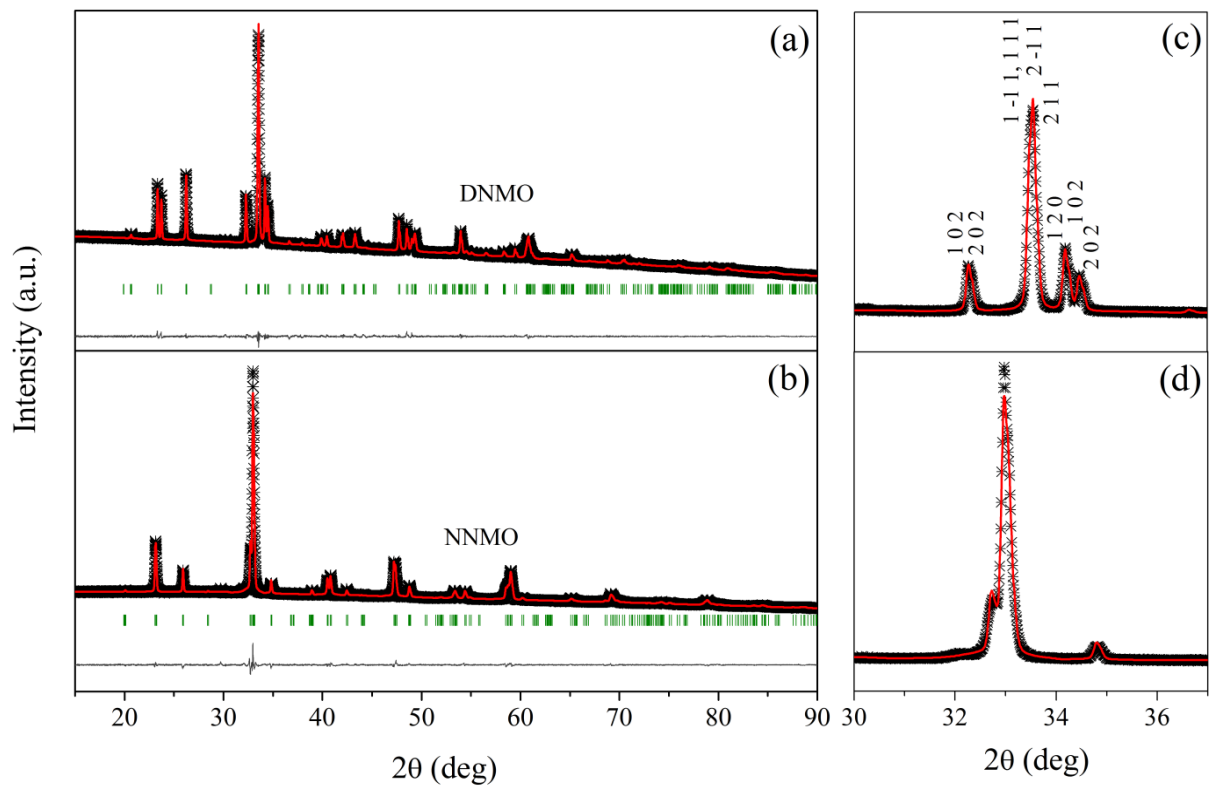

Fig. 2.

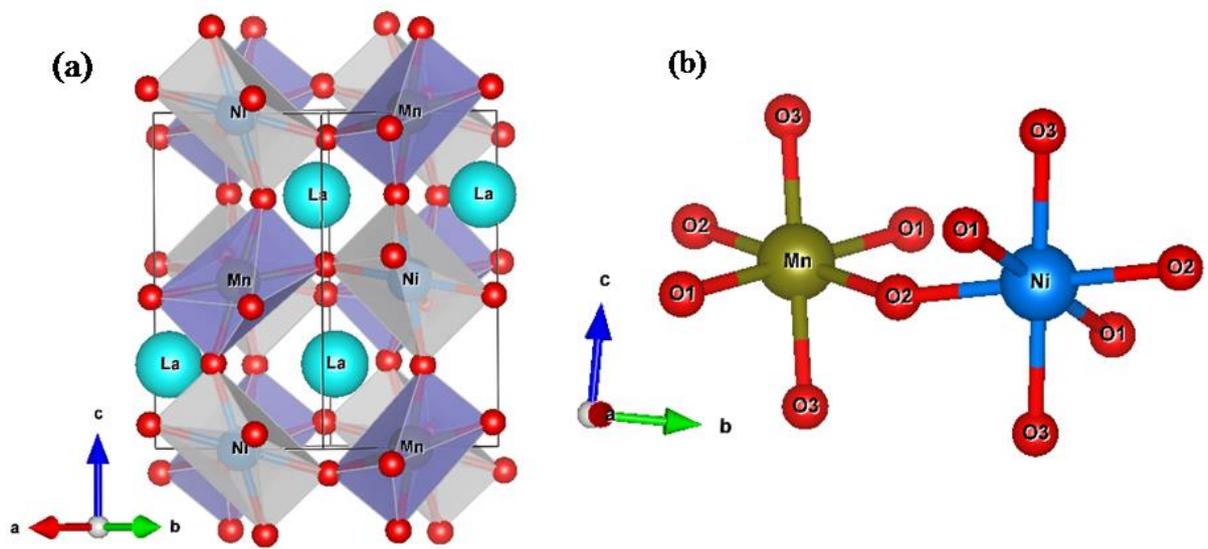

Fig. 3.

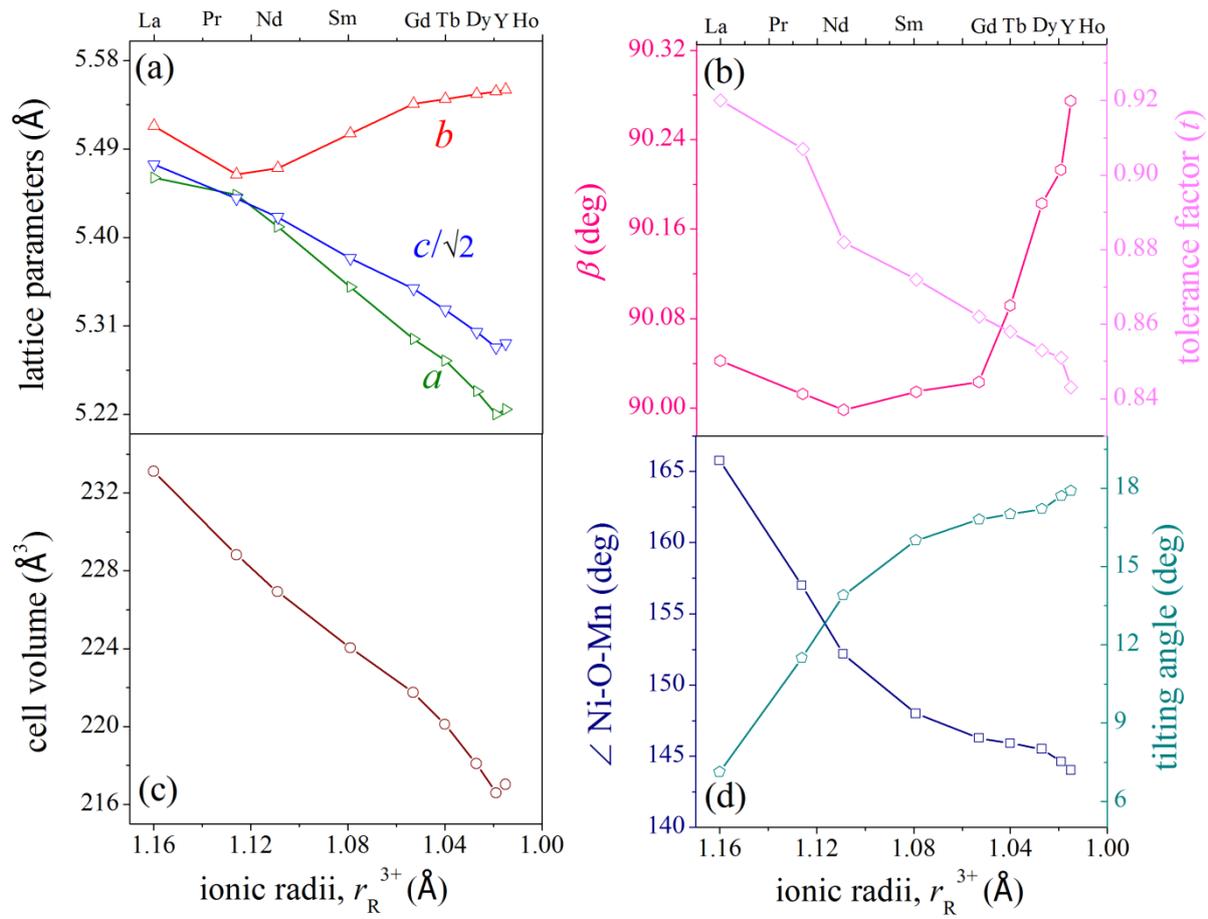

Fig. 4.

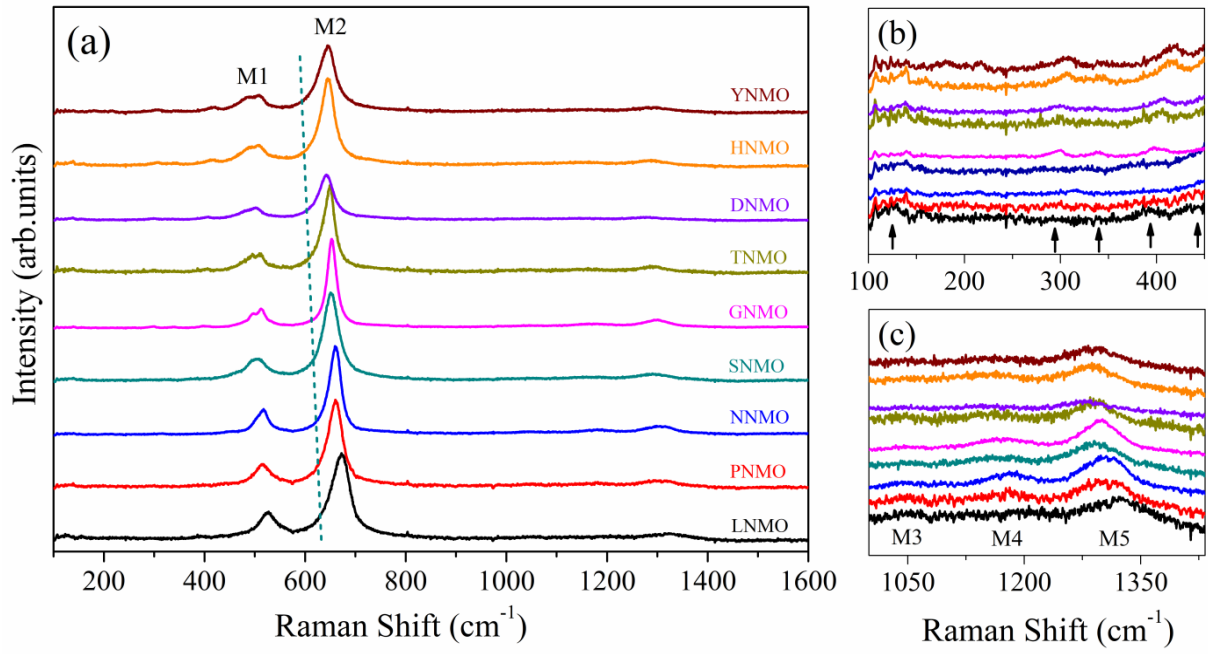

Fig. 5.

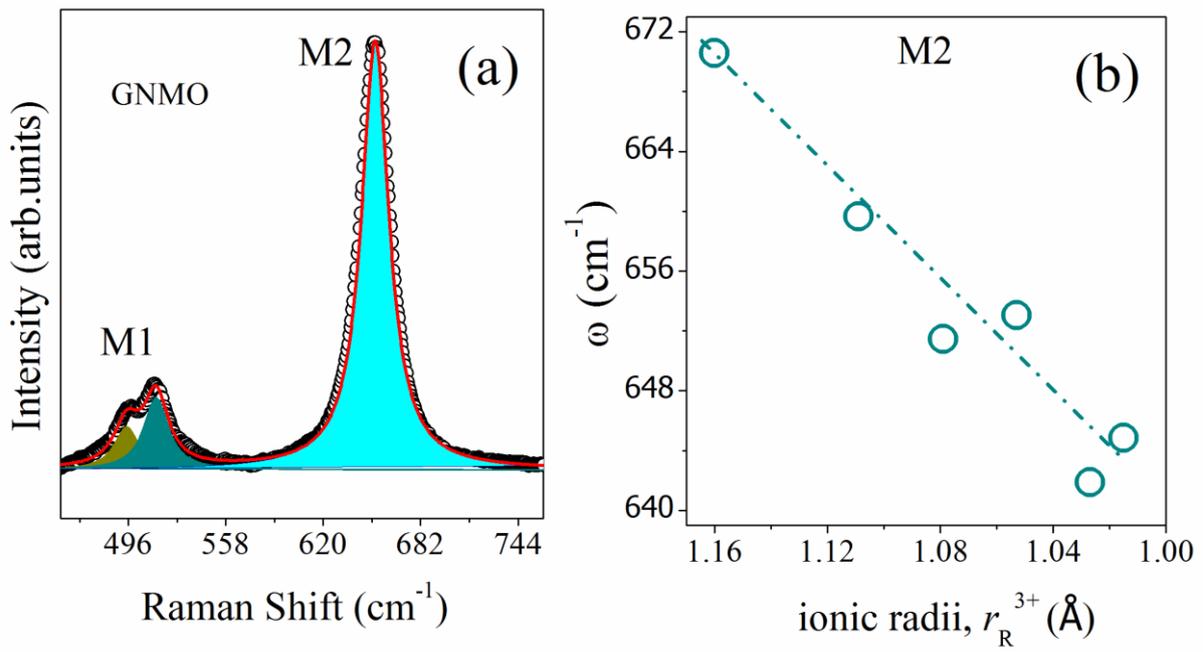

Fig. 6.

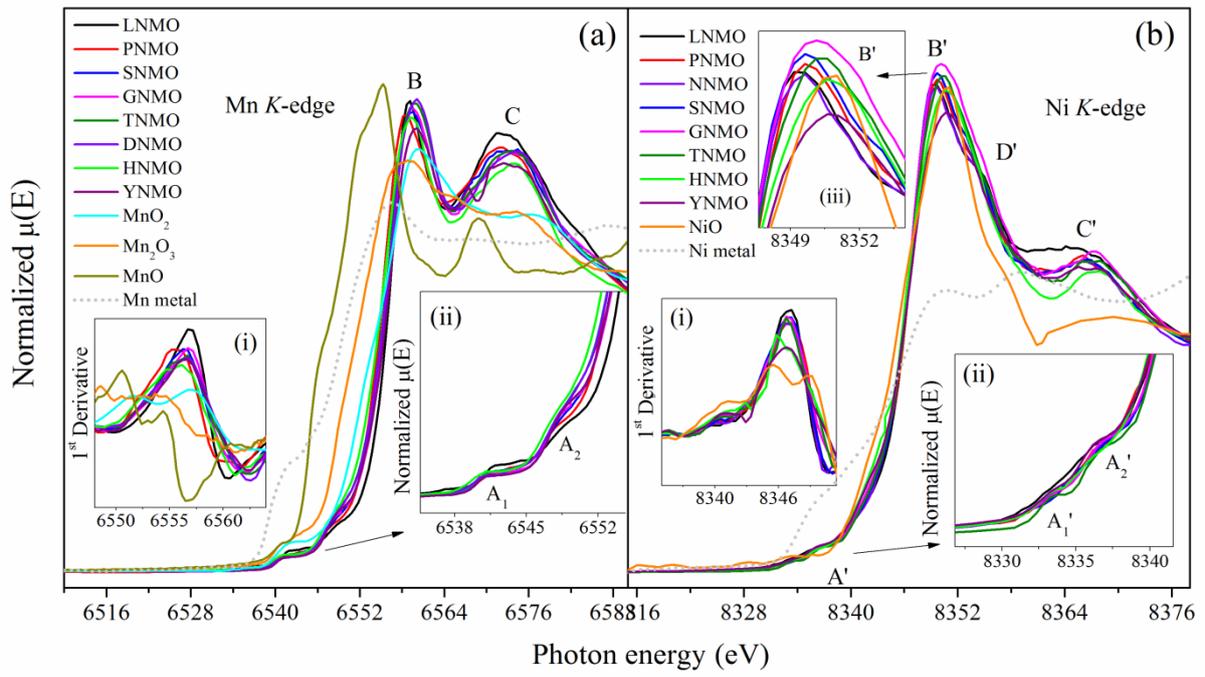

Fig. 7.

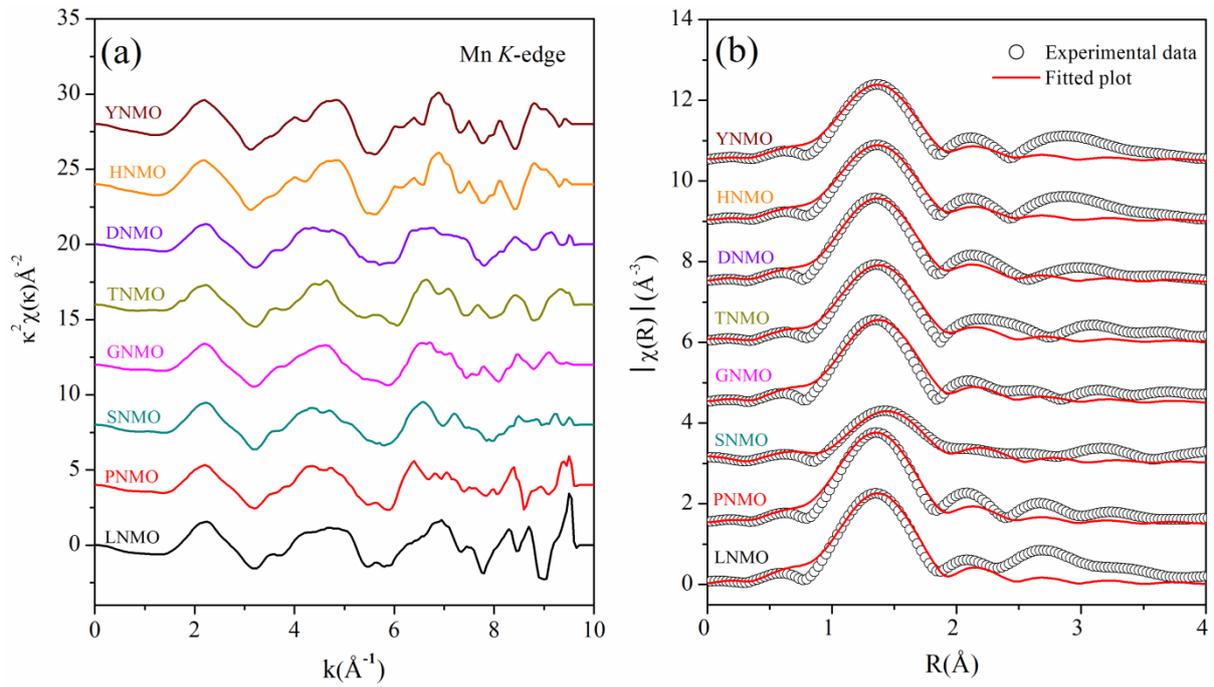

Fig. 8.

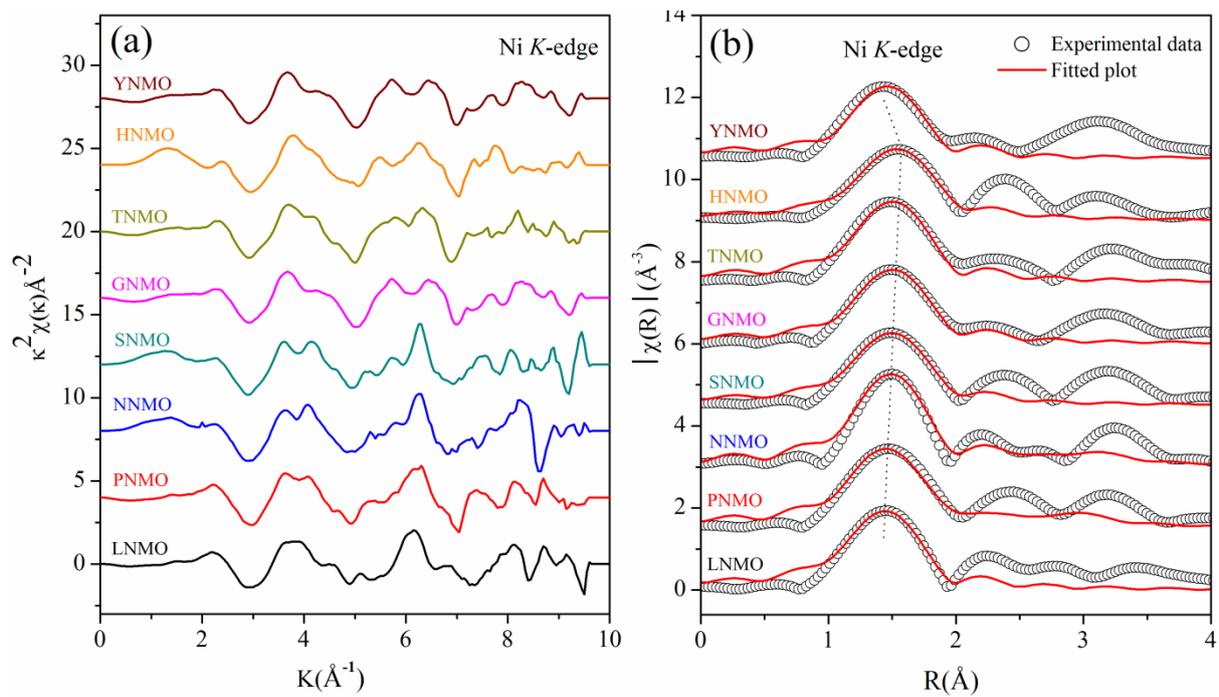

Fig. 9.

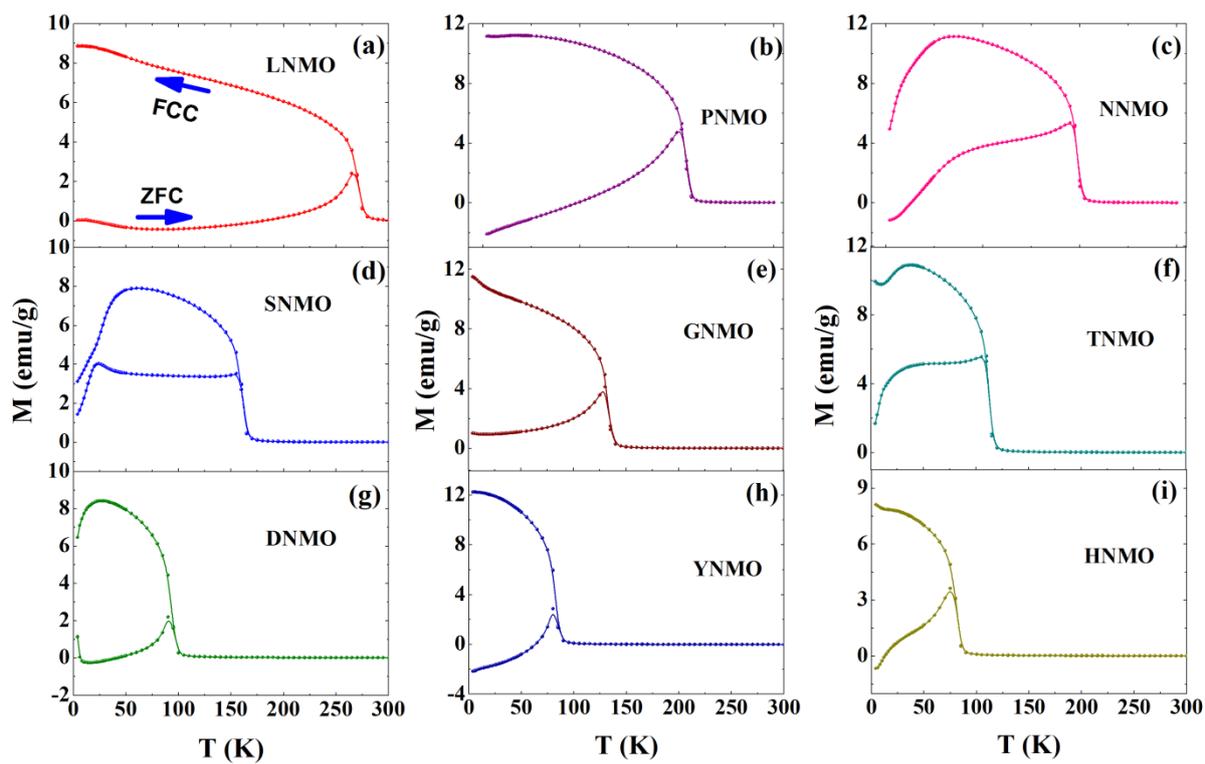

Fig. 10.

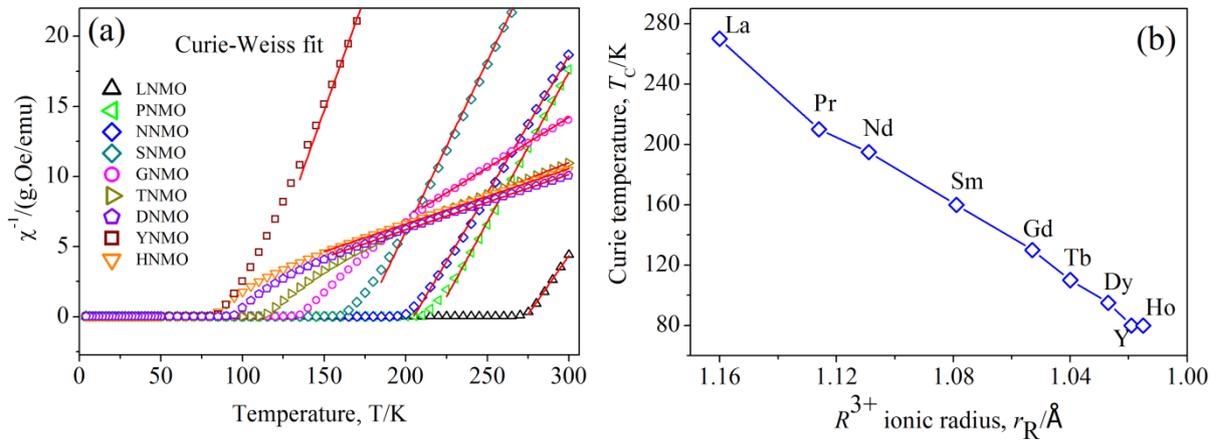

Fig. 11.

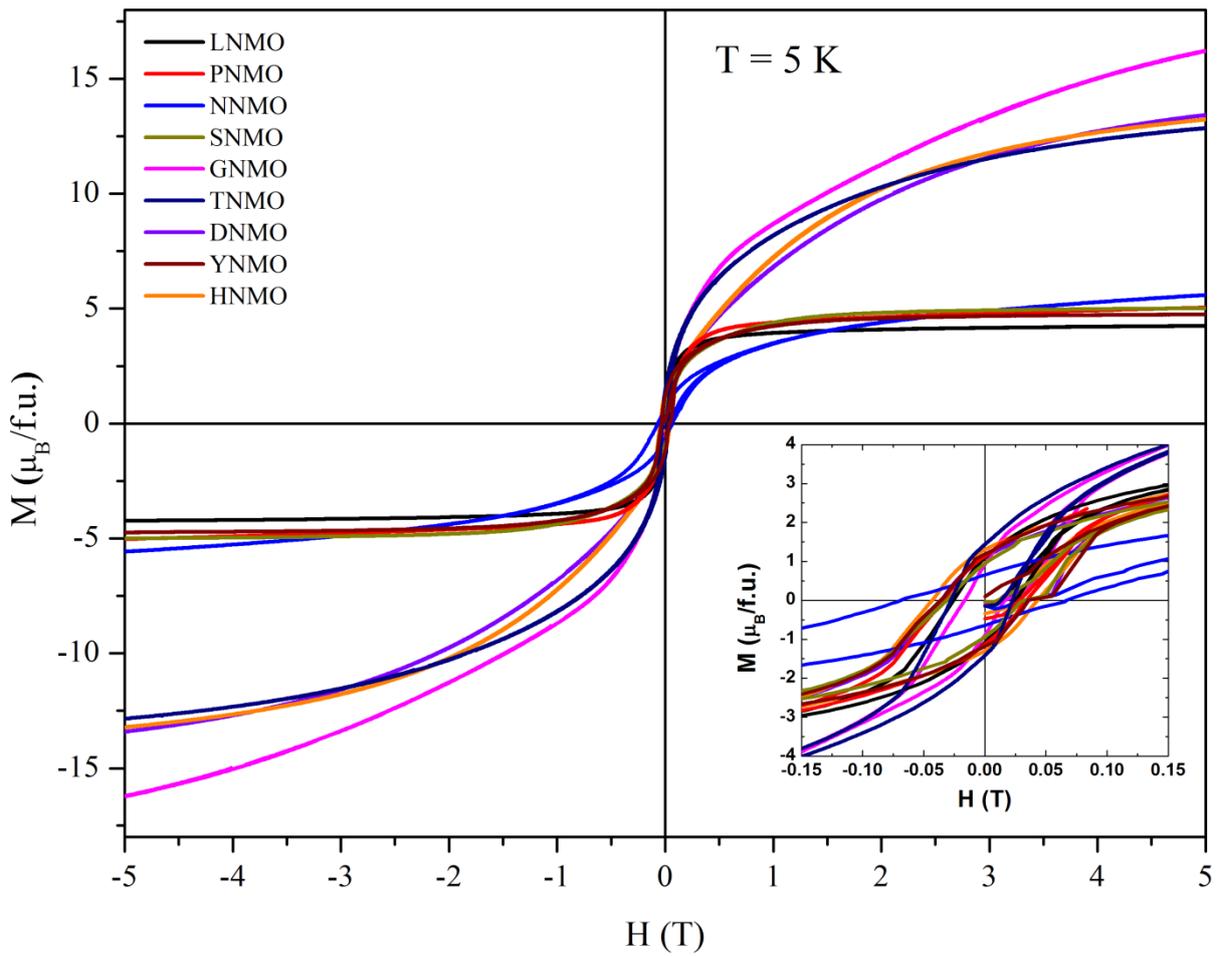

Fig. 12.

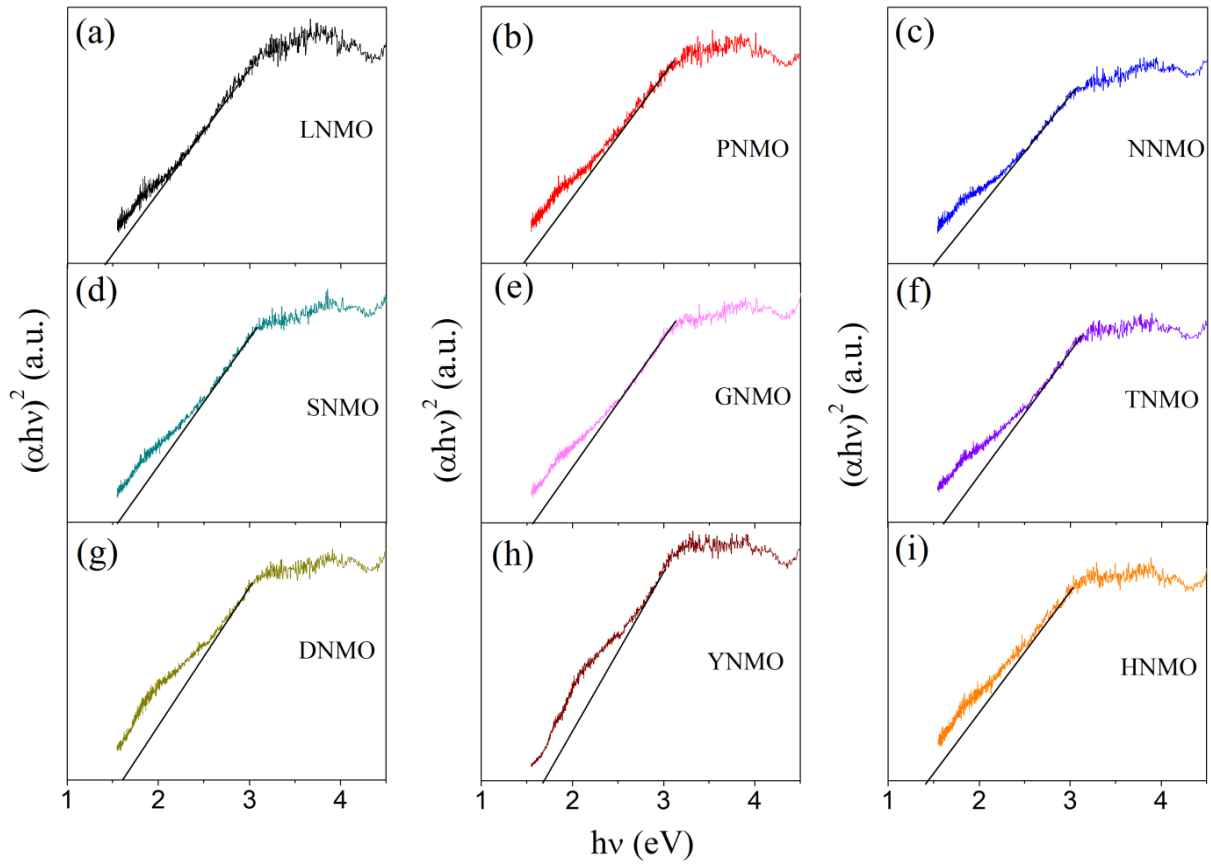

Fig. 13.

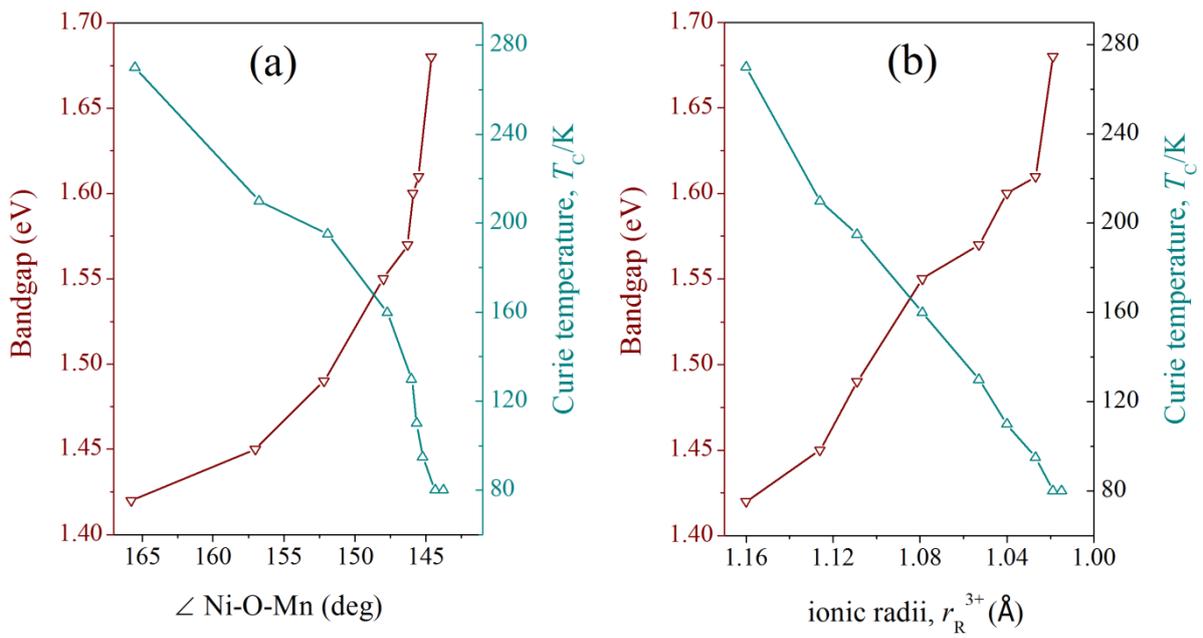

Fig. 14.